\documentclass[10pt,sort&compress]{elsarticle}

\makeatletter
  \long\def\pprintMaketitle{\clearpage
  \iflongmktitle\if@twocolumn\let\columnwidth=\textwidth\fi\fi
  \resetTitleCounters
  \def\baselinestretch{1}%
  \printFirstPageNotes
  \begin{center}%
 \thispagestyle{pprintTitle}%
   \def\baselinestretch{1}%
    {\large\bf\@title}\par\vskip5pt
    \normalsize\elsauthors\par\vskip5pt
    \footnotesize\itshape\elsaddress\par\vskip10pt
    \end{center}%
  \gdef\thefootnote{\arabic{footnote}}%
  }
\makeatother

\makeatletter
\AtBeginDocument{%
  \def\ps@pprintTitle{%
    \let\@oddhead\@empty
    \let\@evenhead\@empty
    \let\@oddfoot\@empty
    \let\@evenfoot\@empty}%
}
\makeatother

\newcommand\blfootnote[1]{%
  \begingroup
  \renewcommand\thefootnote{}\footnote{#1}%
  \addtocounter{footnote}{-1}%
  \endgroup
}

\usepackage[T1]{fontenc}
\usepackage{abstract}
\usepackage[margin=1in]{geometry}
\usepackage{amsmath,amssymb}
\usepackage{booktabs}
\usepackage{array}
\usepackage{listings}
\usepackage{xcolor}
\usepackage{graphicx}
\usepackage{float}
\usepackage{etoolbox}
\usepackage[parfill]{parskip}
\usepackage{url}

\AtBeginEnvironment{tabular}{\footnotesize}

\usepackage[labelfont=bf,font=footnotesize]{caption}
\usepackage[subrefformat=parens,labelformat=parens,labelsep=space,labelfont=small,font=small]{subcaption}
\usepackage{microtype}
\usepackage{textcomp}
\usepackage{pifont}

\usepackage{siunitx}
\DeclareSIUnit{\flop}{FLOP}

\definecolor{lightblue}{rgb}{0.63, 0.74, 0.78}
\definecolor{seagreen}{rgb}{0.18, 0.42, 0.41}
\definecolor{orange}{rgb}{0.85, 0.55, 0.13}
\definecolor{silver}{rgb}{0.69, 0.67, 0.66}
\definecolor{rust}{rgb}{0.72, 0.26, 0.06}
\definecolor{purp}{RGB}{68, 14, 156}
\colorlet{darkrust}{rust!85!black}
\colorlet{darkseagreen}{seagreen!85!black}
\colorlet{darksilver}{silver!65!black}
\colorlet{darklightblue}{lightblue!65!black}

\usepackage{tikz}
\usetikzlibrary{shapes.geometric, arrows.meta, positioning, fit, backgrounds}

\usepackage{hyperref}
\hypersetup{
  colorlinks=true,
}
\usepackage[nameinlink]{cleveref}
\crefname{lstlisting}{listing}{listings}
\Crefname{lstlisting}{Listing}{Listings}

\newcommand\yes{\ding{51}}
\newcommand\no{--}

\definecolor{lstbg}{rgb}{0.965, 0.975, 0.973}   
\journal{}

\begin{document}

\hypersetup{
  linkcolor=darkrust,
  citecolor=seagreen,
  urlcolor=darkrust,
  pdfauthor={Spencer H. Bryngelson},
}

\begin{frontmatter}

  \title{\large\bfseries ANEForge: Python for direct computation on the Apple Neural Engine}

  \author{Spencer~H.~Bryngelson}

  \address{Walker Department of Mechanical Engineering, The University of Texas at Austin, Austin, TX 78712, USA\vspace{-0.35cm}}
  \address{Oden Institute for Computational Engineering and Sciences, The University of Texas at Austin, Austin, TX 78712, USA}

  \date{}

\end{frontmatter}

\blfootnote{
\noindent Email: \url{shb@utexas.edu}}

\vspace{-4ex}
\begin{abstract}
The Apple Neural Engine (ANE) is the fixed-function neural accelerator on every recent Apple device, and the only sanctioned route to it is CoreML.
CoreML treats the engine as a scheduling option: no configuration requires it, the CPU remains a permitted fallback, and no runtime interface reports which unit ran a call.
We present ANEForge, a Python package that compiles a lazy tensor graph into one ANE program and dispatches it through Espresso, the private runtime that CoreML is built on.
It needs no entitlement and leaves system-integrity protection on.
The dispatch unit is fixed, so a measurement attributes to known silicon.
The package reaches the engine's native fused-attention layer, streams int8, int4, and sparse weights, keeps decoder and optimizer state resident across calls, runs the forward pass, backward pass, and optimizer update of training on the engine, and compiles fixed-iteration Krylov solvers and a factored Fourier transform as static-dataflow programs.
On an M5 Pro, a pretrained ResNet-18 forward runs in \qty{0.33}{\milli\second} and uses \qty{2.3}{\milli\joule} per inference, about a factor of 9 less than the GPU at matched fp16 precision (paired runs give \num{5.5} to \num{12.7}).
A call costs about \qty{74}{\micro\second} whether the program fuses one operator or 128.
ResNet-18, a Vision Transformer, and a sentence encoder each compile to one program and match their framework references at cosine similarity \num{1.0000}.
The Stable Diffusion U-Net and VAE decoder also compile to one program apiece, and a single forward pass agrees with the reference to a relative error of \num{0.041} for the U-Net and \num{0.021} for the decoder.
\end{abstract}

\vspace{0.5ex}

\section{Introduction}
\label{sec:intro}

The Apple Neural Engine is a fixed-function neural accelerator that ships on every recent iPhone, iPad, and Apple Silicon Mac~\cite{hollemans_neuralengine, apple2022anetransformers}.
On the M-series Mac it draws the least power of the system-on-chip's three programmable compute blocks for the workloads it accepts.
The only sanctioned route to it is CoreML, which compiles a model and selects a compute unit at runtime~\cite{apple_coreml}.

CoreML treats the engine as a scheduling option.
A caller can exclude the GPU and can inspect placement offline, per layer in Xcode's performance reports or as the estimate \texttt{MLComputePlan} returns.
However, no configuration requires the ANE, the CPU remains a permitted fallback, and no runtime interface reports which unit executed a given call~\cite{apple_coreml, hollemans_neuralengine}.
A workload that the engine accepts and runs efficiently can therefore be served on a slower, less efficient unit with no diagnostic at the call site.

The hardware is reached through a stack of private, undocumented Apple frameworks that CoreML, MPSGraph, and Espresso call internally (\cref{fig:stack}).
That stack is the programming surface of the engine, and it is not exposed to application code.
A study of what the ANE can compute, and a frontend that targets it directly, require working at this level.

\begin{figure}[t]
  \centering
  \begin{tikzpicture}[
    font=\small,
    layer/.style={draw, rounded corners=2pt, minimum height=5.5mm, align=center, inner sep=4pt},
    private/.style={layer, fill=silver!16, draw=darksilver, minimum width=58mm},
    public/.style={layer, fill=lightblue!25, draw=darklightblue, minimum width=26mm},
    afbox/.style={layer, fill=seagreen!10, draw=darkseagreen, minimum width=26mm},
    unit/.style={layer, fill=silver!10, draw=darksilver, minimum width=13mm},
    anebox/.style={layer, fill=rust!12, draw=darkrust, minimum width=13mm},
    fallback/.style={-latex, darksilver, dashed},
    afflow/.style={-latex, darkrust, very thick},
    pubflow/.style={-latex, darklightblue, very thick},
    lbl/.style={font=\small\itshape, text=darksilver},
    x=1mm, y=1mm,
  ]
    \node[private] (espresso) at (0, 0)   {Espresso runtime: \texttt{e5rt} C interface};
    \node[private] (aned)     at (0, -8) {\texttt{aned} daemon: compiles/signs program};
    \node[private] (driver)   at (0, -16) {ANE kernel driver};
    \begin{scope}[on background layer]
      \node[draw=darksilver, dashed, rounded corners=3pt, fill=silver!6,
            inner sep=1.5mm, fit=(espresso) (aned) (driver),
            label={[lbl, anchor=north east, yshift=2pt]south east:private frameworks}] (bg) {};
    \end{scope}

    \node[anebox] (ane) at (0, -27)   {Apple Neural Engine (ANE)};
    \node[unit]   (gpu) at (-38, -27) {GPU};
    \node[unit]   (cpu) at (-57, -27) {CPU};

    \node[public] (coreml) at (-18, 13) {CoreML};
    \node[afbox]  (forge)  at (18, 13)  {\bf ANEForge};

    \draw[pubflow] (coreml.south) -- ([xshift=-12mm]espresso.north)
        node[lbl, pos=0.45, left=2pt] {\normalfont\color{seagreen}if scheduled};
    \draw[afflow] (forge.south) -- ([xshift=12mm]espresso.north)
        node[lbl, pos=0.45, right=2pt, text=darkrust] {\normalfont{}always};
    \draw[fallback] (coreml.west) to[out=180, in=90]
      node[lbl, pos=0.55, left=2pt] {permitted fallbacks} (cpu.north);
    \draw[fallback] ([xshift=-2mm]coreml.west) to[out=190, in=90] (gpu.north);
    \draw[pubflow, darksilver,-stealth] (espresso) -- (aned);
    \draw[pubflow, darksilver,-stealth] (aned) -- (driver);
    \draw[pubflow, darksilver,-stealth] (driver) -- (ane);
  \end{tikzpicture}
  \caption{
  Where ANEForge enters the system.
  CoreML schedules a compiled model across the CPU, the GPU, and the engine, and may serve any call on a fallback unit (dashed); ANEForge enters the private stack at the \texttt{e5rt} interface and dispatches to the engine only (solid).
  Either way, every program is compiled and signed by Apple's \texttt{aned}.}
  \label{fig:stack}
\end{figure}

Here, we present ANEForge, a Python package that compiles an operator graph and dispatches it to the ANE through those private symbols, without loading CoreML.
The caller builds a graph, compiles it to one ANE program, and runs it on the engine alone.
The package exposes its dispatch paths and a machine-readable record of its operators with the evidence that each runs on the hardware.

\Cref{sec:prior} places the package among the projects that reach the engine below CoreML and states what distinguishes its entry point.
\Cref{sec:design} describes the access path, the operator surface, and the compile-and-dispatch pipeline, and \cref{sec:functionality} lists what the package does.
\Cref{sec:evaluation} reports latency, energy, and agreement against reference frameworks on an M5 Pro, with the dispatch floor and the attention rewrite treated in detail, and \cref{sec:stability} reports how the private surface has behaved across operating-system releases.
\Cref{sec:limitations} states the limitations and \cref{sec:conclusions} concludes.

\section{Prior direct access to the engine}
\label{sec:prior}

\Cref{tab:prior} compares the earlier projects with ANEForge.
\texttt{tinygrad} dispatched single operators to the engine from macOS in 2020--2021 by extracting compiled \texttt{hwx} programs and submitting them through the ANE services framework and its IOKit interface, but only by signing a restricted Apple entitlement into the Python interpreter after patching the entitlement check~\cite{hotz_tinygrad_ane}.
The Asahi Linux effort documents the hardware and drives it on Linux through an open kernel driver and a user-space library; its compiler converts programs that macOS's own ANE compiler produced~\cite{yoon_ane_asahi}.
A community knowledge base documents the engine as observed through CoreML~\cite{hollemans_neuralengine}.
Two 2026 projects program the M4 engine through the private \texttt{\_ANEClient} and \texttt{\_ANECompiler} classes of the AppleNeuralEngine framework: a reverse-engineering series with a training repository in Objective-C~\cite{singh2026anem4}, and Orion, an Objective-C runtime with a 27-operation graph intermediate representation that lowers through five passes to MIL and trains a 110M-parameter transformer with the forward and backward passes on the engine and the loss, weight gradients, and Adam update on the CPU~\cite{kumaresan2026orion}.

\begin{table}[t]
  \centering
  \caption{Projects that reach the ANE below CoreML, by what they enter through and what they run.
  ``Entry'' is the private surface the project calls; ``Unentitled'' means no Apple entitlement and no change to code signing or system-integrity protection.
  ``Training'' lists which of the forward pass (F), backward pass (B), and optimizer update (O) run on the engine.
  ``Not stated'' marks a feature the project's documentation does not claim; a dash marks a feature it does not have.
  $^{a}$Inferred from the classes each project calls.
  Singh and Orion use the in-memory-model classes of the AppleNeuralEngine framework, which the companion guide finds callable without an entitlement~\cite{bryngelson2026ane}, but neither project states its entitlement status.}
  \label{tab:prior}
  \setlength{\tabcolsep}{3pt}
  \begin{tabular}{>{\raggedright\arraybackslash}p{2.2cm} >{\raggedright\arraybackslash}p{2.4cm} >{\raggedright\arraybackslash}p{2.4cm} >{\raggedright\arraybackslash}p{2.3cm} >{\raggedright\arraybackslash}p{2.4cm} >{\raggedright\arraybackslash}p{2.8cm}}
    \toprule
    & \texttt{tinygrad} (2021)~\cite{hotz_tinygrad_ane} & Asahi \texttt{ane}, \texttt{anecc}~\cite{yoon_ane_asahi} & Singh (2026)~\cite{singh2026anem4} & Orion (2026)~\cite{kumaresan2026orion} & ANEForge \\
    \midrule
    Entry surface & \texttt{hwx} via ANE services and IOKit & Linux driver; \texttt{hwx} compiled on macOS & \texttt{\_ANEClient}, \texttt{\_ANECompiler} & \texttt{\_ANEClient}, \texttt{\_ANECompiler} & \texttt{e5rt} (Espresso); in-memory model for bridge sub-programs \\
    Unentitled & \no{} (entitlement, patched \texttt{amfid}) & Linux & \yes$^{a}$ & \yes$^{a}$ & \yes \\
    Language & Python & C, Python & Objective-C & Objective-C & Python; Objective-C++ shim \\
    Graph compiler & \no{} (single ops) & \no{} (CoreML compiles) & hand-written MIL kernels & 27-op IR, five passes & lazy graph; 100 fused ops, 19 native layers \\
    Training on engine & \no & \no & F, B & F, B (O on CPU) & F, B, O \\
    Native fused attention & \no & \no & \no{} (decomposed) & \no{} (decomposed) & \yes \\
    Compressed weights & \no & \no & int8 (W8A8) & not stated & int8, int4, sparse \\
    Resident state & \no & \no & not stated & KV cache on host & KV cache, optimizer state \\
    Numerical kernels & \no & \no & \no & \no & Krylov, FFT, Cholesky, eigenvalue \\
    Models validated & none end to end & CoreML models & Stories110M, Qwen3-0.6B & GPT-2 124M, Stories110M & ResNet-18, ViT-B/16, MiniLM, SD U-Net and VAE \\
    License & MIT & MIT & MIT & MIT & MIT \\
    \bottomrule
  \end{tabular}
\end{table}

ANEForge enters the private stack one layer above these projects.
CoreML sits on the Espresso runtime, whose execution runtime \texttt{e5rt} owns compilation, the program library, the execution stream, and the operand descriptors, and \texttt{e5rt} in turn is a client of the AppleNeuralEngine framework's \texttt{\_ANEClient} classes, which hand programs to the \texttt{aned} daemon and the kernel driver~\cite{bryngelson2026ane}.
CoreML is one client of that runtime; an application that names the engine as its target is another.
Both entries converge on \texttt{aned}, which compiles, signs, and caches every program, and reach one engine.
The \texttt{e5rt} compile call accepts MIL with the options Espresso passes and returns a program that \texttt{aned} has lowered and signed, so the fused route gets Apple's operator fusion and layout.
Its ports bind to buffer objects, and binding an output port to an input port's buffer keeps a decoder's key-value cache or a trainer's optimizer state on the engine between calls.
The in-memory models of \texttt{\_ANEClient} are compiled by the same \texttt{aned}, and its requests take IOSurface buffers.
Whether a caller can reuse those buffers across calls to keep state on the engine has not been tested~\cite{bryngelson2026ane}, so Orion's host-side key-value cache (\cref{tab:prior}) may be a design choice of that project.
At \texttt{e5rt} the package reaches compilation and buffer binding through the C interface CoreML uses.
The \texttt{\_ANEClient} classes are entered only for the native bridge layers of \cref{sec:surface}, whose network description the public compiler never emits, and there ANEForge uses the in-memory-model surface that Orion and Singh use.
Unlike the other projects in \cref{tab:prior}, ANEForge keeps a capability registry that pairs its operators with evidence from the hardware (\cref{sec:surface}).

\section{Design}
\label{sec:design}

The package is organized around one path: a lazy operator graph is lowered to a single ANE program and dispatched to the hardware.
This section describes the access path that carries the dispatch (\cref{sec:access}), the operator surface the caller builds graphs from (\cref{sec:surface}), the compile-and-dispatch pipeline (\cref{sec:pipeline}), and the cost model, optimizer, and autograd modules that operate over it (\cref{sec:support}).

\subsection{The access path}
\label{sec:access}

The caller constructs a graph of \texttt{Tensor} nodes from the operator surface.
The compiler lowers the whole graph to the Model Intermediate Language (MIL) that the ANE compiler consumes, packs the weights into one binary blob, and produces a callable program handle.
Dispatch goes through \texttt{e5rt}, a C interface exported by Apple's private Espresso framework, into the \texttt{aned} system daemon and the ANE kernel driver that Apple's own frameworks use.
The program handle is reused across calls, so a model compiles once and runs many times.

A small Objective-C++ shim carries the dispatch; it is compiled on the user's machine because it links private Apple frameworks, and the Python package loads it lazily on the first call.
Access is unentitled.
The shim resolves the symbols of the private frameworks at run time, carries no Apple entitlement, and leaves code signing and system-integrity protection in place.
Every program binary is produced and signed by Apple's own \texttt{aned}, and the shim loads nothing that \texttt{aned} did not sign.
The shim uses about 30 of the 235 \texttt{e5rt} symbols Espresso exports, and the system log shows each dispatch reaching the driver with the code-signing identity of a third-party application.

\subsection{The operator surface}
\label{sec:surface}

We call the \emph{operator surface} the set of operators a caller can build a graph from.
It is a subset of what the hardware accepts, and the capability registry records both sets.
The evidence for the MIL operators is a sweep of all 166 operators through this path, stored in the repository, and the bridge operators have recorded runs of their authored layers~\cite{bryngelson2026ane}.
The registry is generated from the frontend's operator tables, and on hardware a release re-runs only the operator corpus of \cref{sec:stability}.

Two routes feed the dispatch path (\cref{fig:surface}).
The fused route lowers 100 MIL operators, listed by category in \cref{tab:ops}, and four attention composites assembled from them (\texttt{sdpa}, \texttt{mha}, \texttt{cross\_attention}, and \texttt{geglu}) into one program.
Those operators are the subset of the 166-operator MIL vocabulary that a sweep of every operator found to compile and run on the engine through this path (116 of 166) and that the frontend exposes.
The bridge route adds 19 native operators that the public toolchain does not emit, including the fused-attention layer, \texttt{argmax}, \texttt{topk}, \texttt{sort}, and several geometry and point-cloud layers, by authoring the compiler's internal network description directly and running the result as a sub-program through a graph cut.
The 19 are promoted from 26 native hardware layer types authored and run this way; those 26 sit among the 45 layer descriptors the ANE compiler carries, each paired with one of its 50 per-layer validator classes~\cite{bryngelson2026ane}.
The remaining 7 authored layers are recorded in the registry with their evidence but not promoted, because the fused route already covers them or they have no inference-time use.
Four bridge operators have an equivalent fused lowering, and the optimizer of \cref{sec:support} picks one of the two per node.
On the fused route, \texttt{sdpa} lowers to the \texttt{sdpa} composite, \texttt{minmax\_norm} to a composite of \texttt{reduce\_min}, \texttt{reduce\_max}, \texttt{sub}, \texttt{adds}, and \texttt{real\_div}, \texttt{lrn} to \texttt{local\_response\_norm}, and \texttt{flatten} to \texttt{reshape}.

\begin{figure}[t]
  \centering
  \begin{tikzpicture}[
    font=\small,
    set/.style={draw, rounded corners=3pt, align=center, inner sep=3pt},
    fusedset/.style={set, draw=darklightblue, fill=lightblue!18},
    bridgeset/.style={set, draw=darkrust, fill=rust!8},
    both/.style={set, draw=darkseagreen, fill=seagreen!12, minimum height=7mm},
    lbl/.style={font=\small\itshape, text=darksilver},
    x=1mm, y=1mm,
  ]
    \node[fusedset, minimum width=62mm, minimum height=42mm, anchor=north] (mil) at (-34, 0) {};
    \node[anchor=north west, font=\small, text=darklightblue] at ([xshift=1mm, yshift=-1mm]mil.north west) {MIL vocabulary: 166 operators};
    \node[fusedset, minimum width=54mm, minimum height=24mm, anchor=south] (reach) at ([yshift=2mm]mil.south) {};
    \node[anchor=north west, font=\small, text=darklightblue] at ([xshift=1mm, yshift=-1mm]reach.north west) {compile and run on the engine: 116};
    \node[fusedset, fill=lightblue!40, minimum width=46mm, minimum height=14mm, anchor=south] (fused) at ([yshift=2mm]reach.south) {\textbf{fused route: 100 operators}\\+ 4 attention composites\\(\cref{tab:ops})};

    \node[bridgeset, minimum width=60mm, minimum height=42mm, anchor=north] (val) at (34, 0) {};
    \node[anchor=north west, font=\small, text=darkrust] at ([xshift=1mm, yshift=-1mm]val.north west) {ANE compiler per-layer validators: 50};
    \node[bridgeset, minimum width=54mm, minimum height=33mm, anchor=south] (desc) at ([yshift=2mm]val.south) {};
    \node[anchor=north west, font=\small, text=darkrust] at ([xshift=1mm, yshift=-1mm]desc.north west) {hardware layer descriptors: 45};
    \node[bridgeset, minimum width=48mm, minimum height=24mm, anchor=south] (native) at ([yshift=2mm]desc.south) {};
    \node[anchor=north west, font=\small, text=darkrust] at ([xshift=1mm, yshift=-1mm]native.north west) {authored and run: 26};
    \node[bridgeset, fill=rust!20, minimum width=38mm, minimum height=14mm, anchor=south] (bridge) at ([yshift=2mm]native.south) {\textbf{bridge route: 19 operators}\\promoted to the frontend\\(7 recorded, not promoted)};

    \node[both, minimum width=40mm] (sel) at (0, -50) {route-selectable bridge operators: 4\\\texttt{sdpa}, \texttt{minmax\_norm}, \texttt{lrn}, \texttt{flatten}};
    \draw[-latex, darkseagreen] (sel.west) to[out=180, in=270] (fused.south);
    \draw[-latex, darkseagreen] (sel.east) to[out=0, in=270] (bridge.south);
    \node[lbl, anchor=north] at ([yshift=-1mm]sel.south) {the caller builds graphs from the two innermost sets; the registry records all of them};
  \end{tikzpicture}
  \caption{
  How the operator counts relate.
  The fused route (left) exposes 100 of the 116 MIL operators that compile and run on the engine, out of a 166-operator vocabulary; the bridge route (right) exposes 19 of the 26 native hardware layers authored below the public compiler, among the 45 layer descriptors and 50 per-layer validators the ANE compiler carries~\cite{bryngelson2026ane}.
  The four operators at the bottom have a lowering on each route.}
  \label{fig:surface}
\end{figure}

\begin{table}[t]
  \centering
  \caption{The fused operator surface, by category, and the native bridge operators.
  Arithmetic, comparison, matrix-multiply, and reduction operators are also reached through the usual Python operators and methods on tensors (\texttt{+}, \texttt{@}, \texttt{.mean}, and so on).
  \texttt{af.tune} can move \texttt{sdpa}, \texttt{minmax\_norm}, \texttt{lrn}, and \texttt{flatten} from the bridge into the fused program (\cref{sec:surface}).}
  \label{tab:ops}
  \begin{tabular}{l >{\raggedright\arraybackslash}p{0.74\linewidth}}
    \toprule
    Category & Operators \\
    \midrule
    Convolution, pooling & \texttt{conv}, \texttt{conv\_transpose}, \texttt{dynamic\_conv}, \texttt{linear\_activation}, \texttt{max\_pool}, \texttt{avg\_pool}, \texttt{upsample} \\
    Matrix multiply & \texttt{matmul}, \texttt{bmm}, \texttt{bmm\_w}, \texttt{einsum} \\
    Activations, rounding & \texttt{relu}, \texttt{relu6}, \texttt{clamped\_relu}, \texttt{leaky\_relu}, \texttt{thresholded\_relu}, \texttt{prelu}, \texttt{elu}, \texttt{gelu}, \texttt{silu}, \texttt{sigmoid}, \texttt{sigmoid\_hard}, \texttt{tanh}, \texttt{scaled\_tanh}, \texttt{softplus}, \texttt{softsign}, \texttt{softmax}, \texttt{clip}, \texttt{threshold}, \texttt{log}, \texttt{exp}, \texttt{exp2}, \texttt{erf}, \texttt{ceil}, \texttt{floor}, \texttt{round}, \texttt{sign} \\
    Elementwise arithmetic & \texttt{add}, \texttt{sub}, \texttt{mul}, \texttt{real\_div}, \texttt{adds}, \texttt{muls}, \texttt{abs}, \texttt{inverse}, \texttt{maximum}, \texttt{minimum}, \texttt{pow}, \texttt{rsqrt}, \texttt{sqrt}, \texttt{square} \\
    Comparison, select & \texttt{equal}, \texttt{not\_equal}, \texttt{greater}, \texttt{greater\_equal}, \texttt{less}, \texttt{less\_equal}, \texttt{logical\_not}, \texttt{select} \\
    Normalization & \texttt{batch\_norm}, \texttt{instance\_norm}, \texttt{layer\_norm}, \texttt{channel\_layer\_norm}, \texttt{rms\_norm}, \texttt{group\_norm}, \texttt{l2\_norm}, \texttt{local\_response\_norm} \\
    Reductions & \texttt{reduce\_sum}, \texttt{reduce\_mean}, \texttt{reduce\_max}, \texttt{reduce\_min}, \texttt{reduce\_l1\_norm}, \texttt{reduce\_log\_sum}, \texttt{reduce\_log\_sum\_exp}, \texttt{reduce\_sum\_square} \\
    Shape, structural & \texttt{reshape}, \texttt{transpose}, \texttt{squeeze}, \texttt{expand\_dims}, \texttt{flatten2d}, \texttt{slice\_by\_size}, \texttt{concat}, \texttt{stack}, \texttt{split}, \texttt{tile}, \texttt{reverse}, \texttt{crop}, \texttt{pixel\_shuffle}, \texttt{pixel\_unshuffle}, \texttt{space\_to\_depth}, \texttt{depth\_to\_space} \\
    Resize, geometry & \texttt{affine}, \texttt{resize\_nearest\_neighbor}, \texttt{resize\_bilinear}, \texttt{upsample\_bilinear} \\
    Trigonometric, dtype & \texttt{sin}, \texttt{cos}, \texttt{atan}, \texttt{cast}, \texttt{const\_array} \\
    Attention (composite) & \texttt{sdpa}, \texttt{mha}, \texttt{cross\_attention}, \texttt{geglu} \\
    \midrule
    Native bridge (graph cut) & \texttt{sdpa}, \texttt{argmax}, \texttt{topk}, \texttt{sort}, \texttt{fps}, \texttt{radius\_search}, \texttt{cross\_product}, \texttt{cross\_correlation}, \texttt{cost\_volume}, \texttt{minmax\_norm}, \texttt{lrn}, \texttt{space\_to\_channel}, \texttt{channel\_to\_space}, \texttt{space\_to\_batch}, \texttt{batch\_to\_space}, \texttt{dynamic\_slice}, \texttt{input\_view}, \texttt{flatten}, \texttt{scaled\_elementwise} \\
    \bottomrule
  \end{tabular}
\end{table}

\subsection{Compilation and dispatch}
\label{sec:pipeline}

\Cref{fig:architecture} shows the path from a graph to the hardware, and \cref{lst:pipeline} writes it as code, one call per stage, with the implementing module named in each comment.
Building a graph runs nothing; \texttt{af.compile} lowers it and packs the weights into one program, and calling that program dispatches it to the engine.
The optimizer and the trainer are entered through one call each, \texttt{af.tune} and \texttt{af.Trainer}, and use this compile-and-dispatch path.
\Cref{fig:eval} traces the listing's first four lines: the lazy graph they build, the single fused program \texttt{af.compile} produces from it, and the one dispatch each call makes.

\begin{figure}[t]
  \centering
  \newcommand{\modline}[1]{{\scriptsize\ttfamily\color{darksilver}#1}}
  \begin{tikzpicture}[
  font=\small,
  box/.style={draw, rounded corners=2pt, minimum height=11mm, minimum width=18mm, align=center, inner sep=4pt},
  stage/.style={box, fill=lightblue!25, draw=darklightblue},
  code/.style={draw, rounded corners=2pt, minimum height=9mm, fill=seagreen!10, draw=darkseagreen, font=\ttfamily\footnotesize, align=left, inner sep=4pt},
  support/.style={draw, rounded corners=2pt, minimum height=9mm, fill=silver!16, draw=darksilver, font=\small, align=center, inner sep=4pt},
  flow/.style={-latex, darkrust, thick},
  drive/.style={-latex, darkseagreen, dashed, shorten >=1pt},
  thin_/.style={-latex, darksilver},
  node distance=11mm and 9mm,
]
    \node[stage] (graph)    {Tensor graph\\[1pt]\modline{graph.py}};
    \node[stage, right=of graph]   (compile)  {lower to MIL,\\pack weights\\[1pt]\modline{\_compile.py}};
    \node[stage, right=of compile] (program)  {one ANE\\program\\[1pt]\modline{\_blob.py}};
    \node[stage, right=of program] (dispatch) {\texttt{e5rt}\\dispatch\\[1pt]\modline{\_runtime.py}};
    \node[stage, right=of dispatch] (silicon) {ANE\\silicon\\[1pt]\modline{aned}};
    \draw[flow] (graph) -- (compile);
    \draw[flow] (compile) -- (program);
    \draw[flow] (program) -- (dispatch);
    \draw[flow] (dispatch) -- (silicon);

    \node[code, above=10mm of graph] (cgraph) {af.input(\dots)\\x.conv()\,@\,W};
    \node[code, above=10mm of compile, xshift=6mm] (ccompile) {af.compile(out,\\\ \ compress=\dots)};
    \node[code, above=10mm of dispatch] (crun) {net(x)};
    \draw[drive] (cgraph) -- (graph);
    \draw[drive] (ccompile.south) -- (compile.north);
    \draw[drive] (crun) -- (dispatch);

    \node[support, below=11mm of compile, xshift=-10mm] (opt) {\texttt{af.tune} optimizer\\\modline{\_cost.py, \_optimize.py}};
    \node[support, below=11mm of program, xshift=10mm] (grad) {\texttt{Trainer} autograd\\\modline{autograd.py}};
    \draw[thin_] (opt) -- (compile);
    \draw[thin_] (grad) -- (program);
  \end{tikzpicture}
  \caption{
  The compilation and dispatch path: the Python the caller writes (top, green) drives each pipeline stage (middle, with the package module that implements it), and the supporting modules (bottom) feed compilation.
  A small Objective-C++ shim (\texttt{\_lib/*.mm}) carries the \texttt{e5rt} dispatch into the \texttt{aned} daemon.}
  \label{fig:architecture}
\end{figure}

\begin{lstlisting}[float=htb, caption={The pipeline of \cref{fig:architecture} in code, one stage per line, with the module that implements each.}, label={lst:pipeline}]

import aneforge as af

x   = af.input((1, 3, 32, 32))          # graph.py    : a lazy Tensor graph node
y   = af.conv(x, W).relu().mean((2, 3)) # graph.py    : ops extend the graph
net = af.compile(y, compress="int8")    # _compile.py : lower to MIL  (+ _blob.py:
                                        #               pack weights -> one blob)
out = net(image)                        # _runtime.py : dispatch to aned / ANE
                                        #               (handle reused across calls)

fast = af.tune(y)                       # _cost.py + _optimize.py : af.tune optimizer
trainer = af.Trainer(loss, params, lr=1e-3, device_optimizer=True)  # autograd.py : Trainer

\end{lstlisting}

\begin{figure}[t]
  \centering
  \begin{tikzpicture}[
    font=\small,
    op/.style={draw=darklightblue, fill=lightblue!25, rounded corners=2pt, minimum height=7mm, inner sep=4pt, font=\small\ttfamily},
    data/.style={draw=darksilver, fill=silver!12, rounded corners=2pt, minimum height=7mm, inner sep=4pt, font=\small\ttfamily},
    anebox/.style={draw=darkrust, fill=rust!12, rounded corners=2pt, minimum height=7mm, inner sep=4pt},
    lbl/.style={font=\small\itshape, text=darksilver},
    edge_/.style={-latex, darksilver, thick},
    x=1mm, y=1mm,
  ]
    \node[op]   (input) at (0, 0)  {input(1,3,32,32)};
    \node[op]   (conv)  at (27, 0) {conv};
    \node[op]   (relu)  at (42, 0) {relu};
    \node[op]   (mean)  at (59, 0) {mean(2,3)};
    \node[data, minimum height=5mm] (W) at (27, 8.5) {W};
    \begin{scope}[on background layer]
      \node[draw=darkseagreen, dashed, rounded corners=3pt, fill=seagreen!1, thick, inner sep=1.5mm,
            fit=(input) (conv) (relu) (mean) (W),
            label={[lbl, text=darkseagreen, anchor=north]south:{\normalfont\texttt{af.compile(y)}: \texttt{W} packed into the program's weight blob}}] (prog) {};
    \end{scope}
    \node[font=\small, text=darkseagreen, anchor=north east]
      at ([xshift=-1mm, yshift=-0.5mm]prog.north east) {one ANE program};

    \node[anebox] (ane) at (95, 0)  {ANE};
    \node[data]   (out) at (115, 0) {out};
    \draw[edge_] (input) -- (conv);
    \draw[edge_] (conv) -- (relu);
    \draw[edge_] (relu) -- (mean);
    \draw[edge_, darkseagreen] (W) -- (conv);
    \draw[-latex, darkrust, very thick] (prog.east |- mean) -- (ane)
      node[lbl, midway, above=0pt, text=darkrust] {\normalfont net(image)};
    \draw[edge_] (ane) -- (out);
  \end{tikzpicture}
  \caption{
  How \cref{lst:pipeline} evaluates.
  \texttt{input}, \texttt{conv}, \texttt{relu}, and \texttt{mean} are the operator nodes of the lazy graph that lines 1--2 build; building runs nothing.
  \texttt{W} is the convolution's weight tensor, packed into the program's weight blob.
  Each program call \texttt{net(image)} is one dispatch to the engine, which returns the output array \texttt{out}.}
  \label{fig:eval}
\end{figure}

\subsection{Cost model, optimizer, and autograd}
\label{sec:support}

Three modules operate over the two routes.
A structural cost model estimates the time of each node from its shape, dtype, and a calibrated per-architecture table, and \cref{sec:tune} measures how well its ranking agrees with the hardware.
The optimizer \texttt{af.tune} uses that estimate to rank the equivalent lowerings of a graph (route flips between the fused and bridge routes, global or per-weight int8, and constant folding), prunes a lossy candidate whose estimate exceeds 1.5 times the best estimate, measures the survivors on the engine within a budget of eight, validates each against the unoptimized baseline, and caches the result keyed by structure and shape.
A reverse-mode autograd module emits ordinary tensor operators for each gradient, so the backward pass and the optimizer update compile to ANE programs and run on the engine alongside the forward pass.

\section{Functionality}
\label{sec:functionality}

The package provides the following capabilities, each tested on Apple Silicon against a NumPy or framework reference.

\textit{Graph compilation and dispatch.}
\texttt{af.compile} lowers a graph to one ANE program and returns a callable; \texttt{af.input} and \texttt{af.image\_input} declare float and uint8 input ports, the latter emitting an on-engine dequantization so raw camera or decoded-video bytes feed the model directly.

\textit{Weight compression.}
\texttt{af.compile(\dots, compress=)} streams int8, int4 lookup-table, or unstructured-sparse weights from the engine's own dequantization path rather than folding them to half precision at compile time; the weight footprint drops by about 4 times for int4 and 2 for int8 (a 4096-square matmul's weight blob shrinks from \qty{33.6}{\mega\byte} to \qty{8.4}{\mega\byte}).
Compression is off by default, and the default path's program is byte-identical to plain half precision.

\textit{Native attention.}
\texttt{af.sdpa} targets the engine's dedicated fused-attention layer, for causal attention, the single-query decode shape over a cached key and value, and a runtime additive mask.
The public toolchain has carried a scaled-dot-product-attention operator since coremltools 8~\cite{apple_coremltools}, but Apple's MIL compiler decomposes it into matmul, softmax, matmul and never emits the native fused layer, which \texttt{af.sdpa} reaches directly.

\textit{Resident on-engine state.}
Two kinds of state stay resident on the engine across steps through output-to-input buffer aliasing: a zero-copy key-value cache for autoregressive decoding, and optimizer state for training.
The host feeds only the new input each step and reads results back at checkpoints.
CoreML reaches key-value-cache residency through its sanctioned stateful-model interface (\texttt{MLState}, macOS 15 and later)~\cite{apple2024llamacoreml}; the buffer aliasing here is lower-level and carries optimizer state over that mechanism.

\textit{On-engine training.}
The autograd module runs the forward pass, the backward pass, and the optimizer update on the engine; a fused softmax-cross-entropy uses the analytic, half-precision-stable gradient, and the differentiable operator set spans the structural and linear-algebra operators, the common activations, the unary math operators, and the layer, RMS, and group normalizations.

\textit{Pretrained model loaders.}
\texttt{af.load} imports a sentence encoder, \texttt{af.load\_resnet18} a torchvision classifier with batch normalization folded at load, and helpers cover a Vision Transformer and a Stable Diffusion U-Net and VAE decoder.
Further loaders cover Llama- and Qwen-family decoders with resident key-value caches (since v0.2.0) and the Whisper encoder and decoder (since v0.4.0).
We do not evaluate these loaders here.

\textit{Numerical and spectral methods.}
\texttt{aneforge.linalg} and \texttt{aneforge.fft} run fixed-iteration Krylov solvers, Cholesky and eigenvalue routines, and a factored fast Fourier transform as static-dataflow graphs, so dense linear algebra and spectral methods run on the engine at a program size independent of the problem dimension.
Unit tests bound the relative error of the Cholesky reconstruction by \num{5e-3} and of the symmetric eigenvalues by \num{1.5e-2} at $n=8$.
We report no other measurement of these two routines.

\textit{Precompile validation.}
\texttt{af.compile(\dots, validate=True)} runs the ANE compiler's own per-layer validators (the 50 of \cref{fig:surface}) on a graph before compiling it, rejecting an invalid shape or dtype much faster than a round-trip compile would.

\section{Evaluation}
\label{sec:evaluation}

This section reports what the package measures on one host.
\Cref{sec:setup} states the configuration and the energy accounting, \cref{sec:models} the pretrained models, \cref{sec:threeunits} ResNet-18 on all three compute units at matched precision, \cref{sec:floor} the dispatch floor and what it bounds, \cref{sec:tune} the attention rewrite and the cost model behind it, \cref{sec:training} training, decode, and numerical kernels, and \cref{sec:crossgen} the rows that were repeated on two older engine generations.

\subsection{Measurement configuration}
\label{sec:setup}

All timings and energies in this section were taken on an Apple M5 Pro under macOS 26.6.2, with ANE compiler 9.509.0 and Espresso 3525.1.1, running ANEForge v0.4.0; each regenerates from a script in the paper's \path{experiments/} directory or from the example that reports it.
Latency is the minimum end-to-end call time over 30 repetitions after warmup, or over 50 for the dispatch-floor probe of \cref{sec:floor}.
Energy per inference is idle-subtracted active package power, the CPU, GPU, and ANE rails as \texttt{powermetrics} reports them at \qty{500}{\milli\second} intervals over a sustained \qty{6}{\second} window, multiplied by the mean per-call time inside that window; the two quantities come from one measurement, and each arm is repeated eight times, interleaved.
GPU baselines are torchvision on Metal Performance Shaders (torch-MPS) in fp32 and fp16 and a graph-compiled MLX port in fp16; CPU baselines are torchvision in fp32, fp16, and bf16.

\subsection{Pretrained models}
\label{sec:models}

The build-then-compile-then-call pattern of \cref{lst:pipeline} runs unchanged on the pretrained networks of \cref{tab:models}.
A pretrained vision model runs in two lines, and one keyword switches its weights to a streamed, on-engine-dequantized encoding (\cref{lst:pretrained}).
The \texttt{compress} argument selects int8, int4 lookup-table, or sparse weights, accuracy-gated against a tolerance and falling back to a coarser encoding or to half precision when a layer does not meet it (\cref{sec:functionality}).

\begin{lstlisting}[float=htb, caption={Pretrained inference and on-engine weight compression.}, label={lst:pretrained}]

import aneforge as af

clf = af.load_resnet18()                 # torchvision weights, BN folded
label = clf(image)[0].argmax()           # one fused ANE program

# 4x smaller weights, streamed and dequantized on the engine, accuracy-gated:
clf4 = af.load_resnet18(compress="int4", compress_atol=0.05)

\end{lstlisting}

\begin{table}[t]
  \centering
  \caption{Pretrained models run end-to-end on the engine, each fused into one program and validated against its reference framework.
  The classifiers and the encoder report cosine similarity of the output against the framework's float32 output and, for the classifiers, label agreement; the two diffusion components validate one forward pass and report relative error, because a full classifier-free-guidance sampling loop does not close in half precision (\cref{sec:limitations}).}
  \label{tab:models}
  \begin{tabular}{l r l l}
    \toprule
    Model & Ops $\to$ program & Agreement vs.\ reference & Reference \\
    \midrule
    ResNet-18            & 49  & cosine \num{1.0000}, top-1 match & torchvision \\
    ViT-B/16             & 288 & cosine \num{1.0000}, top-5 5/5   & torchvision \\
    MiniLM (L6) encoder  & 144 & cosine \num{1.0000}              & \texttt{transformers} \\
    \midrule
    Stable Diffusion U-Net (one forward) & 308 & relerr \num{0.041}   & \texttt{diffusers} \\
    Stable Diffusion VAE decoder (one forward) & 72  & relerr \num{0.021}   & \texttt{diffusers} \\
    \bottomrule
  \end{tabular}
\end{table}

\subsection{ResNet-18 on three units at matched precision}
\label{sec:threeunits}

\Cref{tab:threeunits} reports the pretrained ResNet-18 forward of \cref{tab:models} on the engine, the GPU, and the CPU at fp32 and fp16, with energy per inference for every arm.
At fp16 the GPU runs the forward in \qty{1.92}{\milli\second} under torch-MPS and \qty{1.83}{\milli\second} under graph-compiled MLX, against \qty{0.334}{\milli\second} on the engine, a latency ratio of about \numrange{5.5}{5.8}; at fp32 torch-MPS is faster than its own fp16 path, \qty{1.66}{\milli\second}, and the ratio is \num{5.0}.
The precision match therefore changes the GPU comparison by less than a factor of 1.2, since ResNet-18 at batch one is latency-bound on the GPU.
On the CPU, PyTorch's fp16 and bf16 convolution paths run the forward in \qty{512}{\milli\second} and \qty{505}{\milli\second}, about eighty times slower than its fp32 path at \qty{5.9}{\milli\second}, so the fp32 row is the fairer CPU baseline.

The engine completes a forward in \qty{2.34}{\milli\joule}, drawing \qty{4.43}{\watt} on the ANE rail and \qty{2.43}{\watt} of host CPU during the dispatch loop, \qty{6.9}{\watt} of package in all; torch-MPS fp16 draws \qty{21.8}{\milli\joule}, torch-MPS fp32 \qty{25.6}{\milli\joule}, compiled MLX \qty{34.5}{\milli\joule} (faster, but at \qty{15}{\watt} on the GPU rail), and the fp32 CPU \qty{106}{\milli\joule}.
The engine thus uses about a factor of 9 less energy than torch-MPS at matched fp16 precision, 15 less than compiled MLX, and 45 less than the CPU at fp32.
These factors divide mean energies, and the torch-MPS fp16 energy, \qty{21.8 \pm 6.1}{\milli\joule}, varies most.
Pairing each engine repetition with the torch-MPS fp16 repetition run next to it gives ratios from \num{5.5} to \num{12.7}, median \num{8.5}, against \numrange{13.4}{16.0} for compiled MLX and \numrange{41.5}{48.5} for the fp32 CPU.
The sustained time that energy uses is within \qty{5}{\percent} of the minimum latency for every arm except torch-MPS, where it is longer by a factor of \num{1.28} at fp16 and \num{1.38} at fp32.

\begin{table}[t]
  \centering
  \caption{ResNet-18 ($1\times3\times224^2$) on the three compute units of an M5 Pro.
  Latency is the minimum over 30 calls.
  The sustained column gives the mean call time inside a \qty{6}{\second} power window, over which idle-subtracted active package power (\qty{0.88}{\watt} idle) is sampled.
  Energy is their product, given as mean and standard deviation over eight interleaved repetitions.
  Relative error is against the torchvision fp32 CPU output, and every arm returns the same top-1 label; the MLX port carries timm weights, so its output is not compared.
  The two slow CPU arms were not sampled for energy.}
  \label{tab:threeunits}
  \setlength{\tabcolsep}{3pt}
  \begin{tabular}{l l r r r r r l}
    \toprule
    Unit & Precision & Latency (\unit{\milli\second}) & Sustained (\unit{\milli\second}) & Power (\unit{\watt}) & Energy (\unit{\milli\joule}/inf) & Energy ratio & Rel.\ error \\
    \midrule
    ANE (ANEForge)      & fp16 & 0.334 & 0.341 & 6.89  & \num{2.34 \pm 0.16}  & 1    & \num{2.3e-3} \\
    GPU (torch-MPS)     & fp16 & 1.922 & 2.467 & 9.13  & \num{21.8 \pm 6.1}   & 9.3  & \num{1.7e-3} \\
    GPU (MLX, compiled) & fp16 & 1.832 & 1.875 & 18.40 & \num{34.5 \pm 1.4}   & 14.7 & -- \\
    GPU (torch-MPS)     & fp32 & 1.658 & 2.295 & 11.65 & \num{25.6 \pm 7.5}   & 10.9 & \num{7.0e-7} \\
    CPU (torch)         & fp32 & 5.934 & 6.205 & 17.12 & \num{106.3 \pm 1.6}  & 45.3 & 0 (reference) \\
    CPU (torch)         & fp16 & 511.7 & --    & --    & --                   & --   & \num{1.8e-3} \\
    CPU (torch)         & bf16 & 505.1 & --    & --    & --                   & --   & \num{1.5e-2} \\
    \bottomrule
  \end{tabular}
\end{table}

\subsection{The dispatch floor}
\label{sec:floor}

\begin{figure}[t]
  \centering
  \includegraphics{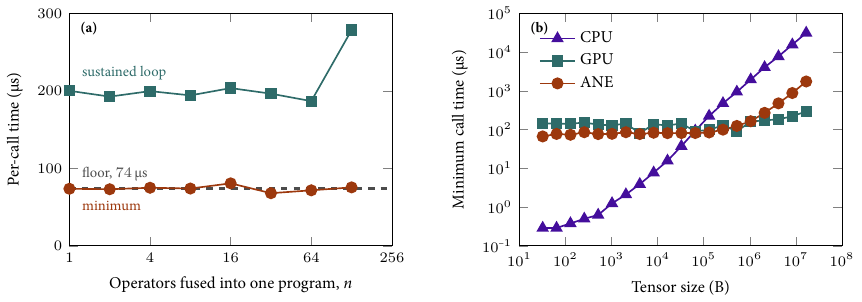}
  \caption{
  The per-program dispatch floor on the M5 Pro.
  (a) Minimum (over 50 calls) and sustained per-call time of a chain of $n$ small \numproduct{1x1} convolutions fused into one program, with the intercept of the linear fit, \qty{74.0}{\micro\second}, dashed.
  (b) Minimum time of one \texttt{relu} over $2^k$ fp16 elements on the three units, as labeled.}
  \label{fig:floor}
\end{figure}

Every call to the engine pays a fixed cost for the host-to-daemon submission and the firmware round trip, whatever the size of the work, and dispatch is single-in-flight, so concurrency does not amortize it.
The size of that floor sets the smallest call for which the engine pays, and it is measured three ways in \cref{fig:floor}.

A chain of $n$ small \numproduct{1x1} convolutions fused into one program runs in about \qty{74}{\micro\second} whether $n$ is 1 or 128; the linear fit has an intercept of \qty{74.0}{\micro\second} and a slope indistinguishable from zero.
The floor is paid once per program, and the 128 operators inside it cost nothing measurable at this size.
A single \texttt{relu} over $2^k$ fp16 elements runs in about \qty{77}{\micro\second} (median) from \qty{32}{\byte} to \qty{128}{\kibi\byte}, after which the host transfer term appears and the latency reaches \qty{1.75}{\milli\second} at \qty{16}{\mebi\byte}, a rate of about \qty{9.6}{\giga\byte\per\second} for the copy into and out of the bound buffers.
In a sustained loop the per-call time is about \num{2.6} times the minimum-latency probe (\qty{195}{\micro\second} against \qty{74}{\micro\second}); the latencies of \cref{tab:threeunits} are probe minima.
The companion guide reports an M1 floor of \qtyrange{0.19}{0.23}{\milli\second} per call~\cite{bryngelson2026ane}.
The lower figure is the mean of a hot loop over a small convolution graph with host copies, and the upper comes from a slope fit.
About \qty{98}{\percent} of the M1 floor is dispatch overhead, \qty{130}{\micro\second} of it the firmware round trip, and a fused stack there stays flat from one layer to thirty-two.

That \texttt{relu} on the other two units sets the crossover.
On the CPU (NumPy) it completes in under \qty{1}{\micro\second} at the smallest sizes and crosses the engine's floor at \qty{64}{\kibi\byte}; on the GPU (MLX) it pays a floor of its own, \qtyrange{130}{150}{\micro\second}, which the engine beats below \qty{512}{\kibi\byte} and loses to above it, because the GPU's transfer term is smaller.
For a single memory-bound operator the engine is thus not the fastest unit at any size tested; the CPU finishes first below \qty{64}{\kibi\byte} and the GPU above \qty{512}{\kibi\byte}, so the engine's advantage comes from fusion.
A program the cost model predicts to be floor-bound should do more work per call, by increasing the batch or by chaining more operators into one program, and the package warns about this at compile time.

\subsection{The attention rewrite and the cost model}
\label{sec:tune}

The optimizer rewrites an attention block by folding its query, key, value, attention, and output projections into one fused program, a transformation verified bit-identical to the unoptimized graph.
\Cref{tab:sweep} reports the rewrite across sequence lengths for an eight-head block with 64-dimensional heads, and for the attention shape of ViT-B/16.
For sequence lengths up to 256 the rewrite runs about \numrange{3.8}{5.3} times faster, because the unrewritten block is five dispatches deep and each pays the floor of \cref{sec:floor}.
At 512 and 1024 the tuner keeps the native route, since the attention itself then accounts for most of the block's cost and the graph cut no longer costs a measurable fraction; the measured speedup of the rewrite is \num{1.00}.
On the ViT-B/16 shape the rewrite runs \num{4.05} times faster.
The whole pass is one call: \texttt{af.tune(out)} prunes the equivalent lowerings of a graph with the cost model, measures the survivors on the engine, caches the result keyed by structure and shape, and validates each against the unoptimized baseline, returning the lossless selection by default.
An explicit tolerance, \texttt{af.tune(out, atol=0.1)}, admits an int8 rewrite only when it also clears a speedup margin.

\begin{table}[t]
  \centering
  \caption{The attention rewrite across sequence lengths $S$ on the M5 Pro.
  Latency is the minimum over 30 calls; the tuned column is the lowering \texttt{af.tune} selected, and ``native'' means it kept the unrewritten graph.
  The tuned output matched the unoptimized output to a relative max-abs difference of \num{0.0000} in every row.}
  \label{tab:sweep}
  \begin{tabular}{r r r r r l}
    \toprule
    Heads & $S$ & Head dim & Unoptimized (\unit{\micro\second}) & Tuned (\unit{\micro\second}) & Speedup \\
    \midrule
    8 & 16   & 64 & 498  & 104  & $4.78\times$ \\
    8 & 32   & 64 & 492  & 92   & $5.34\times$ \\
    8 & 64   & 64 & 503  & 104  & $4.82\times$ \\
    8 & 128  & 64 & 615  & 125  & $4.93\times$ \\
    8 & 256  & 64 & 844  & 220  & $3.83\times$ \\
    8 & 512  & 64 & 605  & 607  & $1.00\times$ (native kept) \\
    8 & 1024 & 64 & 1746 & 1747 & $1.00\times$ (native kept) \\
    12 & 197 & 64 & 917  & 226  & $4.05\times$ \\
    \bottomrule
  \end{tabular}
\end{table}

\Cref{tab:pruning} tests the cost model as a pruning heuristic.
For each benchmark graph, every legal lowering was compiled and measured with no pruning and no budget, and the tuner's rule was then replayed against the full table.
The variant space is small, one to three lowerings per graph, because the legal variants are the route flips and the global int8 rewrite.
In the three attention graphs the rule pruned the global-int8 variant, and the measurement confirms it was slower than the route flip the model ranked first; in no graph was the measured optimum pruned, and the tuner's selection was the measured optimum in every case.
The estimate is coarse in absolute terms since it carries no dispatch term; the rewritten attention block, for example, is estimated at \qty{63}{\micro\second} and measures \qtyrange{106}{227}{\micro\second}.
The model's ranking agreed with measurement in every graph with more than one variant.
The calibrated table is one curve per ANE architecture, fourteen at this release (H11 through H17s), interpolated from recorded sweeps, so on another chip the model uses that chip's curve and the M5 Pro table is not applied elsewhere.
Whether that curve ranks as well on another chip was not measured.

\begin{table}[t]
  \centering
  \caption{The cost model as a pruning heuristic on the M5 Pro.
  Every legal lowering of each graph was measured; the tuner's rule (prune a lossy candidate estimated at more than 1.5 times the best estimate, measure at most eight) was then replayed against the measurements.}
  \label{tab:pruning}
  \begin{tabular}{l r r l l}
    \toprule
    Graph & Variants & Pruned & Optimum pruned & Tuner's choice \\
    \midrule
    ResNet-18                        & 2 & 0 & no & fp16 (measured optimum) \\
    MiniLM encoder ($S{=}32$)        & 2 & 0 & no & fp16 (int8 variant rejected on accuracy) \\
    Attention, 8 heads, $S{=}32$     & 3 & 1 & no & fused rewrite (measured optimum) \\
    Attention, 8 heads, $S{=}256$    & 3 & 1 & no & fused rewrite (measured optimum) \\
    Attention, 12 heads, $S{=}197$   & 3 & 1 & no & fused rewrite (measured optimum) \\
    CIFAR-10 CNN ($B{=}128$)         & 1 & 0 & no & fp16 (only variant) \\
    \bottomrule
  \end{tabular}
\end{table}

\subsection{Training, decode, and numerical kernels}
\label{sec:training}

\Cref{lst:train} trains a convolutional network on the engine.
The convolution is assembled from primitives so its weight is a graph parameter, and the group normalization is built from primitives so its affine trains at any batch size; the forward pass, the backward pass, and the Adam update all run on the engine.
The network reaches \qty{71.2}{\percent} test accuracy on CIFAR-10, within \num{1.7} points of a PyTorch model of the same layer topology at \qty{72.9}{\percent}; the convolution and group-normalization gradients match PyTorch to 1.0 cosine similarity.

\begin{lstlisting}[float=htb, caption={Train a convolutional network forward, backward, and optimizer step on the engine.}, label={lst:train}]

import aneforge as af

x, logits, params = af.cifar_cnn(batch=128)        # conv -> GroupNorm -> relu -> pool
target = af.input((128, 10))
obj = af.softmax_cross_entropy(logits, target)     # fp16-stable analytic gradient
tr = af.Trainer(obj, params, lr=3e-3, optimizer="adam",   # fwd + bwd + Adam on ANE
                device_optimizer=True, data_inputs=first_batch)
for images, labels in batches:                     # host feeds only data + lr
    tr.data[x], tr.data[target] = images, onehot(labels)
    tr.step()                                      # one on-engine training step
acc = tr.accuracy(test_images, test_labels)        # 71.2% (PyTorch peer: 72.9%)

\end{lstlisting}

The single-query decode shape lets a decoder run autoregressively on the engine (\cref{lst:decode}): one new query attends over a cached key and value.
The shipped example \texttt{examples/gpt\_generate\_ane.py} wraps this primitive in a full decode loop whose key-value cache stays resident on the engine across steps.
The host feeds only the new token and a position mask, the cache never returns to the host, and the output matches a NumPy reference token for token.

\begin{lstlisting}[float=htb, caption={Single-query decode-shape attention, the public primitive behind on-engine autoregressive decode.}, label={lst:decode}]

import aneforge as af

# one new query (seq_q = 1) attends over the cached keys and values of length S
a = af.sdpa(q.reshape(1, H, 1, dh),       # the new token's query
            Kc.reshape(1, H, S, dh),      # cached keys
            Vc.reshape(1, H, S, dh))      # cached values
# examples/gpt_generate_ane.py wraps this in a full loop whose KV cache stays
# resident on the ANE across steps (share_buffer); host feeds only the new
# token + a position one-hot, and the output matches numpy token for token.

\end{lstlisting}

The engine has no data-dependent control flow, but a fixed-iteration numerical method is static dataflow and compiles to one program whose size does not depend on the problem dimension.
The package runs a fixed-iteration Krylov solve for symmetric positive-definite and general systems on the engine, and a fast Fourier transform staged as a product of factors and carried as real and imaginary pairs (\cref{lst:scientific}).
The test suite checks both against NumPy in double precision.
The solve must reach a relative error of \num{5e-3} at $n=48$ and condition number 10 (\num{5e-2} at condition number 100), and the two-dimensional transform \num{2e-2} at sizes up to $64\times64$.

\begin{lstlisting}[float=htb, caption={Dense linear algebra and spectral methods on the engine.}, label={lst:scientific}]

from aneforge.linalg import conjugate_gradient
from aneforge.fft import fft as ane_fft

x = conjugate_gradient(A, b, iters=40)   # fixed-iteration CG; same program at any n
Xre, Xim = ane_fft(sig_re, sig_im, N)    # factored DFT; matches np.fft to fp16

\end{lstlisting}

The repository's \texttt{examples/} directory contains runnable versions of these demonstrations, and the operator corpus is the project's correctness gate.
The operator counts and reference agreements of \cref{tab:models} are emitted by the corresponding scripts (\path{examples/resnet18.py}, \path{vit.py}, \path{sentence_embeddings.py}, \path{sd_unet.py}, \path{sd_vae.py}), the attention-rewrite speedups and the pruning audit by \path{experiments/attention_sweep.py} and \path{experiments/cost_model_pruning.py}, and the three-unit and dispatch-floor measurements by \path{experiments/resnet18_three_units.py} and \path{experiments/dispatch_floor.py}, so each number regenerates on Apple Silicon.

\subsection{Across engine generations}
\label{sec:crossgen}

The timings above come from one M5 Pro, and the package is verified on an M1 Max only by the operator corpus.
The repository's single-stream device map, a fixed set of synthetic workloads the bench tooling runs on each host (\path{bench/device_compare_wattcomplete.py}), was however run on an M1 Max, an M2 Pro, and the M5 Pro under one harness, with the results committed as \path{bench/results/device_compare_wattcomplete_results_M{1,2,5}.json}, and \cref{tab:crossgen} reproduces the rows that bear on this paper.
On the M1 Max, M2 Pro, and M5 Pro, the floor row takes \qtylist{164;110;92}{\micro\second}, the ResNet-like convolution stack \qtylist{1.92;1.27;0.76}{\milli\second}, and the ViT attention shape \qtylist{0.61;0.59;0.23}{\milli\second}.
In the same harness the M2 Pro runs the pretrained ResNet-18 forward in \qty{1.06}{\milli\second} at \qty{2.6}{\milli\joule} per inference, and the M5 Pro in \qty{0.33}{\milli\second} at \qty{2.2}{\milli\joule}.
The map ranks the units the same way for each workload on all three hosts, with multipliers that change by generation.
From the M1 Max to the M5 Pro the floor row falls by a factor of \num{1.8} and the convolution stack by a factor of \num{2.5}.
Apple lists a 16-core Neural Engine for both chips.
The ANE compiler, which assigns output channels across engine units, plans for 8 units on the M1 Max target (H13G) and 16 on the M5 Pro target (H17s), with cost-model clocks of \qty{1.14}{\giga\hertz} and \qty{1.89}{\giga\hertz}~\cite{bryngelson2026ane}.
The floor row times one $64\times256\times256$ matmul that moves \qty{32}{\kibi\byte} each way, against \qty{2}{\kibi\byte} for the convolution chain of \cref{sec:floor}, and on the M5 Pro the \texttt{relu} probe puts the cost of that larger transfer at about \qty{8}{\micro\second}.
The other \qty{10}{\micro\second} of the gap between \qty{74}{\micro\second} and \qty{92}{\micro\second} was not isolated.
The timings of \cref{sec:threeunits,sec:floor,sec:tune} are thus single-host results, and \cref{tab:crossgen} is the extent of what was repeated elsewhere.

\begin{table}[t]
  \centering
  \caption{Rows of the repository's single-stream device map repeated on three engine generations under one harness (the script and the committed result files are named in \cref{sec:crossgen}).
  Latency is the minimum over 30 calls on the engine (15 for ResNet-18), and energy per inference follows \cref{sec:setup} over a single \qty{6}{\second} window.
  The rows were recorded in June 2026 with a pre-release version of the package.
  \Cref{tab:threeunits} reports \qty{2.34}{\milli\joule} for ResNet-18, an average of eight windows under v0.4.0.}
  \label{tab:crossgen}
  \begin{tabular}{l r r r}
    \toprule
    Workload & M1 Max & M2 Pro & M5 Pro \\
    \midrule
    GEMM floor ($M{=}64$, $K{=}256$, $N{=}256$), \unit{\milli\second} & 0.164 & 0.110 & 0.092 \\
    Convolution stack ($C{=}256$, $32^2$, $3\times3$, depth 16), \unit{\milli\second} & 1.918 & 1.265 & 0.757 \\
    Attention ($S{=}197$, $D{=}768$, 12 heads), \unit{\milli\second} & 0.613 & 0.589 & 0.230 \\
    Layer norm ($197\times768$), \unit{\milli\second} & 0.321 & 0.283 & 0.195 \\
    ResNet-18 forward, \unit{\milli\second} & -- & 1.064 & 0.330 \\
    ResNet-18 forward, \unit{\milli\joule}/inf & -- & 2.64 & 2.20 \\
    \bottomrule
  \end{tabular}
\end{table}

\section{Stability of the private surface}
\label{sec:stability}

The package calls private, undocumented Apple symbols that carry no API contract and can change between operating-system releases.
This section reports what has been observed across releases.

Each release records the macOS and ANE-compiler versions it was verified against, and the operator corpus is run on the engine before a release ships.
The corpus has 80 small graphs of one to ten operators, and ten of them are also compiled with int8 weights, for 90 variants in all.
A variant passes when it compiles, runs on the engine, and matches a NumPy reference.
The graphs use 67 of the 100 fused operators and the bridge operators \texttt{sdpa} and \texttt{argmax}.
Separate on-engine tests compare the two lowerings of \texttt{sdpa}, \texttt{minmax\_norm}, \texttt{lrn}, and \texttt{flatten}.
Dispatch is verified on the M5 Pro and the M1 Max under macOS 26.5 (ANE compiler 9.509.0, Espresso 3520.4.1) and on the M5 Pro under macOS 26.6.2, where the Espresso component had moved to 3525.1.1 and the compiler had not, with the operator corpus (90 of 90 variants) and every measurement of \cref{sec:evaluation} passing on the newer release.
The package requires macOS 14 or later, and on macOS 14 and 15 the verification is narrower.
Hosted continuous-integration runners virtualize macOS without exposing the engine, so since July 2026 a weekly scheduled job builds the dispatch shim, imports the package, and runs every hardware-free test on macOS 14, macOS 15, and the newest hosted image, and every one of its six runs to date has passed on all three.
That job shows the shim still links and the package still imports on those releases; it does not dispatch, and the corpus gate on macOS 14 or 15 hardware has not been run.

Across the releases the package has run on, only one dispatch symbol changed.
macOS 26 introduced \path{e5rt_execution_stream_submit_async} and deprecated the earlier asynchronous submit, and the shim resolves whichever is present.
The compiler and the silicon have broken more often than the symbols.
A differential fuzzer that compiles random graphs on the engine and compares them against NumPy found six cases in which Espresso accepts a program and miscompiles it, each now carried as a workaround in the lowering with a regression test.
Five further behaviors differ silently between hosts or releases and are recorded as pitfalls with their probes: a slice kernel keyed to the compiler target that differs between M1 and M5, saturation of slices above 4094 elements on M1, dynamic-weight convolution at batch two or more, a compile-service restart of about \qty{15}{\second}, and four-character-code image input~\cite{bryngelson2026ane}.
The package is for research and local development, since software linking private frameworks cannot ship through the App Store.

\section{Limitations}
\label{sec:limitations}

The package presented here has limitations.
The route is interoperability research on hardware the user owns, not a circumvention.
The package carries no Apple entitlement, leaves code signing and system-integrity protection in place, and redistributes no Apple material, with the dispatch shim compiled on the user's machine.
An archived source capsule accordingly preserves the source, the corpus, and its expected outputs.
Executability requires an Apple Silicon machine within the verified range, because the shim links the operating system's resident private frameworks and the program binary comes from its ANE compiler.

The compute path is half precision; accumulation error stays acceptable for the validated models, but half precision sets the dataplane's numerical reach.
A full classifier-free-guidance diffusion sampling loop, for example, does not close in half precision: the guided difference it amplifies is a fraction of a percent of the signal, below the half-precision step error.

On the machines tested the engine exposes a single hardware lane, so a host dispatch drives one program at a time and the throughput gains come from fusing work into one program, not from concurrent dispatch.
ANEForge is not a general tensor library.
A graph that uses an operator outside the operator surface fails at compile time.
The timings of \cref{sec:evaluation} are single-host results, and \cref{sec:crossgen} bounds what was repeated on other generations.

\section{Conclusions}
\label{sec:conclusions}

ANEForge gives Python direct access to the Apple Neural Engine through Espresso, with the compute unit fixed and no entitlement.
The engine's operator census is narrower than its compiler's vocabulary, 116 of 166 MIL operators and 26 native layers below the public compiler, and the per-program floor of about \qty{74}{\micro\second} sets the smallest unit of work for which a dispatch pays.
A single memory-bound operator finishes sooner on the CPU below \qty{64}{\kibi\byte} and on the GPU above \qty{512}{\kibi\byte}, so the engine's advantage comes from fusion; a program of 128 operators pays the floor once.
On ResNet-18 the engine uses about a factor of 9 less energy per inference than the GPU at matched precision, with paired runs from \num{5.5} to \num{12.7}.
A fixed-iteration method compiles to the engine if its program size is independent of the problem dimension and it uses only operators from the operator surface.
Dense Krylov solvers and a factored Fourier transform qualify, and half precision limits how ill-conditioned a problem they can solve.

\section*{Declaration of competing interest}

The author declares no competing financial interests or personal relationships that could have influenced the work reported in this paper.

\section*{Data availability}

ANEForge v0.4.0 is available under the MIT license at \url{https://github.com/sbryngelson/ANEForge}, with documentation at \url{https://aneforge.readthedocs.io}.
The version this paper describes is archived at \url{https://doi.org/10.5281/zenodo.20672609}.
The package is written in Python and Objective-C++ and requires an Apple Silicon Mac running macOS~14 or later, Python~3.10 or later, and NumPy; ANE dispatch is verified on macOS 26 (\cref{sec:stability}).

\end{document}